\documentclass[intlimits,twoside,a4paper]{article}

\usepackage[cp1251]{inputenc}
\usepackage[eqsecnum]{cmpj3}

\usepackage{bm}

\issue{2022}{25}{1}{13501}
\doinumber{10.5488/CMP.25.13501}
\title[Equilibrium properties of the two-level lattice fluid]%
{Equilibrium properties of the lattice fluid with the repulsion between the nearest neighbors on the
two-level lattice with nonrectangular geometry%
}
  \author[Ya. G. Groda]{Ya. G. Groda\orcid{0000-0003-4470-8388}}
  \address{Belarusian State Technological University, 13a Sverdlova str., 220006 Minsk, Belarus}

\Keywords{lattice fluid, two-level lattice, quasi-chemical approximation, 
	diagram approximation, Monte Carlo simulation}

\date{Received August 6, 2021, in final form December 6, 2021}
\begin{document}

\maketitle

\begin{abstract}
The equilibrium properties of the lattice fluid with the repulsion between the nearest neighbors
on the two-level planar triangular lattice are investigated. The numerical results obtained 
from the analytical expressions are compared with the Monte
Carlo simulation data. It is shown that the previously proposed diagrammatic
approximation makes it possible to determine the equilibrium characteristics of the lattice fluid with the
repulsion between the nearest neighbors on a two-level lattice with an accuracy comparable 
to the accuracy of
modelling the system using the Monte Carlo method in the entire range of thermodynamic parameters. It was
found that, in contrast to a similar one-level system, a lattice fluid with the repulsion between the
nearest
neighbors undergoes a first-order phase transition. 
%
%
\printkeywords
\end{abstract}

\section*{Introduction}

The model of the lattice gas or the lattice liquid is widely used in the description of processes
occurring on the surfaces and in the volumes of solids \cite{Domb}, 
as well as in the study of various electrochemical systems~\cite{Strom, Vakarin, Bisquert1, Levi,
Bisquert2}.

The main idea of this model consists in the decomposition of the original physical system into two spatial
interpenetrating subsystems, one of which is rather rigidly structured and plays the role of the reference
one, and in the second one the particles retain high mobility. In fact, the particles of the second subsystem move in
the field created by the supporting subsystem. In this case, it is assumed that the given field
does not depend on the distribution of the moving particles. Thus, it can be argued that the reference
subsystem creates a potential with a spatially periodic profile for the moving particles, in which there are
sufficiently deep potential energy minima that form one or other periodic lattice.

Another condition for the possibility of using the lattice fluid model to describe a real physical system
is the presence of two significantly different time scales. The first of which, $\tau_1$, is determined by
the average residence time of the mobile particle in the minimum of potential energy, and the second,
$\tau_2$, is determined by the oscillation period of the mobile particle near this position. 
Hence, for the indicated time scales, we have $\tau_1 \gg \tau_2$.

The fulfillment of the above conditions, the stability of the reference system and the presence of two
significantly different time scales, allows us to restrict ourselves to considering only the subsystem of
moving particles, and the characteristics of the potential landscape can be considered as the input
parameter of the model, assuming them to be given. 

The potential energy minima 
will be determined as the lattice sites, each of which can be either occupied
by a single particle or can be vacant. When considering a system of charged particles occupying the lattice sites,
the number of possible states of each lattice site obviously increases \cite{Groda21} in comparison with
the case of uncharged particles. In the case of a system with charged particles,  each lattice site can be
occupied by a positive particle, a negative particle, or can be vacant.

One of the ways for the development of lattice models is the transition to the lattice models with
energetically nonequivalent lattice sites. For example, the so-called two-level system in which the nodes of
two different types are present can be considered.

Earlier in the literature, several versions of two-level models of the lattice fluid were presented
\cite{Vakarin, Bisquert1, Levi, Vorot, Choj97, Choj99, Choj20, Tar01}. For example, the models with a random
distribution of energetically deeper and shallower nodes were proposed \cite{Vakarin, Bisquert1, Vorot,
Choj99}, as well as the models, in which the nodes of different types form some symmetrical structures
\cite{Vakarin, Choj97, Choj20, Tar01}. In the latter case, it was shown \cite{Levi, Bisquert2} that the model of
a lattice fluid on a two-dimensional two-level lattice can be used to study layered intercalation
compounds, for example, graphite intercalated with lithium ions. In case of such systems, the typical values of deep 
level energies are of the order of several tenths of eV, while shallow level energies can be considerably 
lower or comparable with the former. In many cases,  interparticle interactions are of the order of several 
hundredths of eV.

In \cite{Tar01, Tar01_ss}, the lattice fluid on the 2D rectangular two-level lattice was investigated
using a decoration-iteration transformation, which made it possible to reduce the problem to considering a lattice
gas on a homogeneous square lattice, and investigating the type of the phase diagram. 
In particular, by means of real-space renormalization group (RSRG) method it was established
\cite{Tar01_ss} that in the considered system with repulsion between the nearest neighbors 
the first-order phase transition takes place. This is a quite unusual phase transition of the first order in a
lattice gas system with mutual repulsion between adparticles. The existence of the 
first-order phase transition was also confirmed with the help of Monte Carlo (MC) simulation 
\cite{Tar01}. Moreover, in \cite{Tar01} the equilibrium properties of the lattice fluids with 
interaction between nearest neighbors on the rectangular two-level lattice were 
investigated with the help of RSRG method and MC simulation.

The diffusion  properties
of this model were studied using the Monte Carlo simulation method
\cite{Tar02,Tar02_2,Tar07,Tar08,Tar09,Tar10,Tar15,Tar15_2}. It can also be noted that the number
of approximate methods belonging to the class of mean field methods were developed to determine the
equilibrium properties of this system \cite{monogr}. Within the framework of the constructed approximate
approaches, the thermodynamic, structural and transport properties of systems with repulsion and 
attraction between nearest neighbors \cite{GrodaSSI} on a square lattice were investigated. 

In this article we consider the possibility of constructing approximate approaches for assessing the
equilibrium properties of the lattice fluid on the lattice with non-rectangular geometry. The
quasi-chemical (QChA) and diagram (DA) approximations will be developed to estimate the equilibrium
properties of the two-level lattice system based on the planar triangular lattice. 
The accuracy of both approximations will be verified by comparing the QChA and DA results 
with the Monte Carlo (MC) simulations data. Furthermore, within the framework of the two approaches, 
mean field approximations and MC, the
equilibrium properties of the lattice fluid with the repulsive interaction between the nearest neighbors
on the two-level lattice will be investigated.

\section{Model}

\begin{figure}[!t]
\centerline{\includegraphics[width=0.65\textwidth]{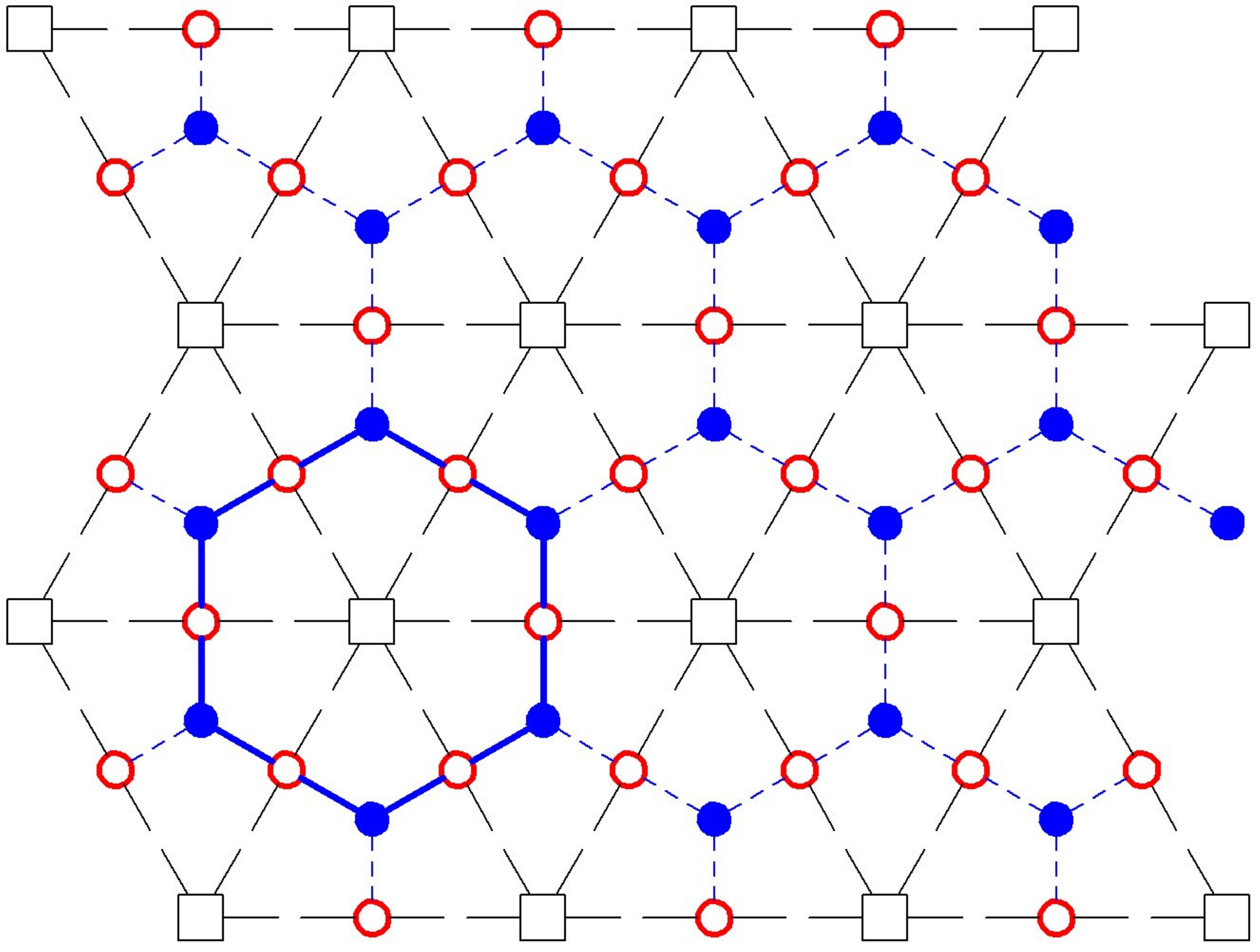}}
\caption{(Colour online) Schematic view of the lattice with two types of sites. The open squares 
correspond to the atoms of the
crystal plane (1, 1, 1), the open and closed circles correspond to the $\alpha$- and $\gamma$-sites of the
lattice model.} \label{fig1}
\end{figure}

The crystal plane (1, 1, 1) of the crystal with the simple cubic Bravais lattice can be considered as the
basis for constructing a two-level lattice system. In this plane, the atoms of the crystal form a
triangular lattice as shown in figure~\ref{fig1}, where these atoms are represented by open squares. 
The interaction
between the atoms of the plane forms an energy profile, the minima of which are the locations of the
particles adsorbed on a given plane of the crystal, the so-called adparticles. The ensemble of these
adparticles forms the system under study.

The geometric features of the original lattice make it possible to distinguish two types of sites of the
lattice model. Some of them --- $\alpha$-sites --- are located between two neighboring atoms of the crystal
plane and
are shown in figure~\ref{fig1} with open circles. The second type of sites --- $\gamma$-sites --- are in the
center of the regular triangle built on three neighboring atoms. 
In figure~\ref{fig1}, $\gamma$-sites are shown as closed circles.

In the general case, the depths of potential wells corresponding to the $\alpha$- and 
$\gamma$-sites can be different 
\begin{equation}
\epsilon_{\alpha}\neq\epsilon_{\gamma}, \quad \delta\epsilon=\epsilon_{\alpha}-\epsilon_{\gamma},
\label{eq1}    
\end{equation}
where $\epsilon_{\alpha}$, $\epsilon_{\gamma}$ denote the depth of the potential well 
corresponding to the lattice sites of the indicated type, and $\delta\epsilon$ 
is the difference in the depths of the potential wells.

Since the $\alpha$ and $\gamma$ sites are not energetically equivalent, the probability of finding the
particle in each
of them can be different. For a theoretical description of this system, the original lattice can be
divided into the system of two sublattices, each of which contains the sites of only one type. This allows us
to speak in what follows about the $\alpha$ and $\gamma$ sublattices, respectively.

Thus, the constructed lattice system is a combination of two two-dimensional hexagonal sublattices. 
In this case, the $\alpha$-sublattice has a lattice constant $a/2$, and the $\gamma$-sublattice has a lattice constant
$a/\sqrt{3}$, where $a$ is the distance between the atoms of the crystal plane.

It can also be noted that the first coordination number, i.e., the number of the nearest neighbor nodes is
different for the sublattices introduced above. Thus, for example, the sublattice site $\alpha$ has
$z_{\alpha}=2$ nearest-neighbor sites. Moreover, all these nearest neighboring sites belong to the
$\gamma$ sublattice. On the other hand, for a $\gamma$-site, the number of nearest sites on the $\alpha$
sublattice is $z_{\gamma}=3$.

In this paper, we consider the lattice fluid with the interaction between nearest neighbors. 
The energy of this interaction is assumed to be equal to $J_1$. 
It should be noted that the particles occupying the lattice sites of the same type do 
not interact with each other, because these nodes are not nearest neighbors.

Counting the sites on each of the sublattices is not difficult. As noted above, the $\alpha$-site is
located between every two nearest atoms of the crystal plane. If we assume that the initial crystal 
plane contains $M$ atoms forming the triangular lattice with the first coordination number $z_M=6$, 
then the number of pairs of nearest atoms, and hence the number of $\alpha$-sites, is
\begin{equation}
N_{\alpha}=\frac{z_M M}{2}.
\label{eq2}    
\end{equation}

The $\gamma$-site lies in turn in the center of each triangular graph built on triples of the 
nearest atoms. The method for calculating the corresponding weight coefficients, meaning 
the number of different graphs of a given type that can be built on the lattice of the selected type,
is described in detail in \cite{monogr}. The weighting coefficient of the three-vertex graph on 
the triangular lattice, and hence the number of $\gamma$-nodes, is
\begin{equation}
N_{\gamma}=2M.
\label{eq3}    
\end{equation}

Thus, in the case of a lattice containing $N$ lattice sites, we obtain 
\begin{equation}
N_{\alpha}=\frac{3}{5}N, \quad N_{\gamma}=\frac{2}{5}N.
\label{eq5}    
\end{equation}

For further calculations, it is convenient to introduce the average first coordination number $z_1$
which has the meaning of the average number of the nearest neighboring sites for an arbitrary 
lattice node. For the lattice under consideration, the coordination number $z_1$ will 
not be an integer. It can be determined by the following relation 
\begin{equation}
z_1N=z_{\alpha}N_{\alpha}+z_{\gamma}N_{\gamma}, \quad z_1=\frac{12}{5}.
\label{eq6}    
\end{equation}

Thus, the original lattice can be considered as the system of two nonequivalent sublattices built on 
the same lattice sites. For each of the sublattices, the concentration of particles $c_1^{\xi}$, 
and vacancies $c_{0}^{\xi}$ can be determined, and the order parameter of the model $\delta c$ 
\begin{equation}
c_{\alpha}=c_{1}^{\alpha}=\frac{n_{\alpha}}{N_{\alpha}}, \quad 
c_{\gamma}=c_{1}^{\gamma}=\frac{n_{\gamma}}{N_{\gamma}}, \quad 
c=c_1=\frac{n_{\alpha}+n_{\gamma}}{N},
\label{eq7}    
\end{equation}
\begin{equation}
\delta c=c_{\gamma}-c_{\alpha}, \quad 
c=\frac{3}{5}c_{\alpha}+\frac{2}{5}c_{\gamma},
\label{eq8}    
\end{equation}
\begin{equation}
c_{\alpha}=c-\frac{2}{5}\delta c, \quad 
c_{\gamma}=c+\frac{3}{5}\delta c,
\label{eq9}    
\end{equation}
\begin{equation}
c_0=1-cz_1, \quad 
c_{0}^{\xi}=1-c_{1}^{\xi}, \quad 
\xi=\alpha,\gamma,
\label{eq10}    
\end{equation}
where $n_{\xi}$ ($\xi=\alpha,\gamma$) is the number of adparticles on the sublattice $\xi$, and $c_i$
is the average concentration of adparticles ($i=1$) and vacancies ($i=0$) in the system.

The thermodynamic state of the constructed lattice model can be defined by specifying the set of 
occupation numbers {$n_i$} taking values $n_i=1$ or $n_i=0$, if the $i$-th lattice site is 
occupied by the particle or is vacant, respectively.

For a given set of occupation numbers, the potential energy of the system of $n$
particles on the lattice can be represented as:
\begin{equation}
U_N=\epsilon_{\alpha}\sum\limits_{\left<\alpha\right>}n_i+
\epsilon_{\gamma}\sum\limits_{\left<\gamma\right>}n_i+
\frac{J_1}{2}\sum\limits_{\left<nn\right>}n_in_j,
\label{eq12}    
\end{equation}
where $\left\langle \alpha\right\rangle $, $\left<\gamma\right>$ and $\left<nn\right>$ denote a summation over all $\alpha$-, $\gamma$-sites 
and over all pairs of nearest-neighbor sites, respectively.

In what follows, an additional condition will be imposed on the considered lattice system
\begin{equation}
\delta\epsilon=\frac{J_1}{2}.
\label{eq13}    
\end{equation}
Compliance with condition (\ref{eq13}) ensures the symmetry of the phase diagram around $c=0.5$.

\section{Monte Carlo simulation algorithm}

Determination of the equilibrium properties of the lattice system can be performed within the 
framework of the standard Metropolis algorithm \cite{metro}, the application of which 
to the lattice fluids of various types is described in detail, for example, in \cite{uebing,CMP,BSU1,BSU2}.

In this case, the modelling of the equilibrium properties of the system is carried out in the 
grand canonical ensemble, i.e., with a variable number of particles in the system. 
Within the framework of this algorithm, the value of the chemical potential $\mu$ 
and the interaction energy $J_1$ is fixed, and the concentration of particles on 
the lattice $c$ and their distribution are determined directly during the simulation. 
In this case, the initial distribution of the particles over the $N$-site lattice can be arbitrary.

After choosing an arbitrary lattice site and changing its state (adding a particle if the site 
is vacant, and removing it if  the site is occupied), the change in the energy $\delta E_{N}$ 
corresponding to the given change in the state of the site is determined as
\begin{equation}
\delta E_N=\pm(J_1s_1-\mu+\epsilon_{\alpha}-\delta\epsilon\delta_{\gamma\xi}),
\label{eq14}    
\end{equation}
where the plus corresponds to the addition of a particle, and the minus to its removal; 
$s_1$ is the number of particles occupying the nearest lattice sites; $\delta_{\gamma\xi}$ 
is the Kronecker symbol; index $\xi$ denotes the type of sublattice to which an arbitrarily 
chosen node belongs. 

\begin{figure}[!t]
\centerline{\includegraphics[width=0.65\textwidth]{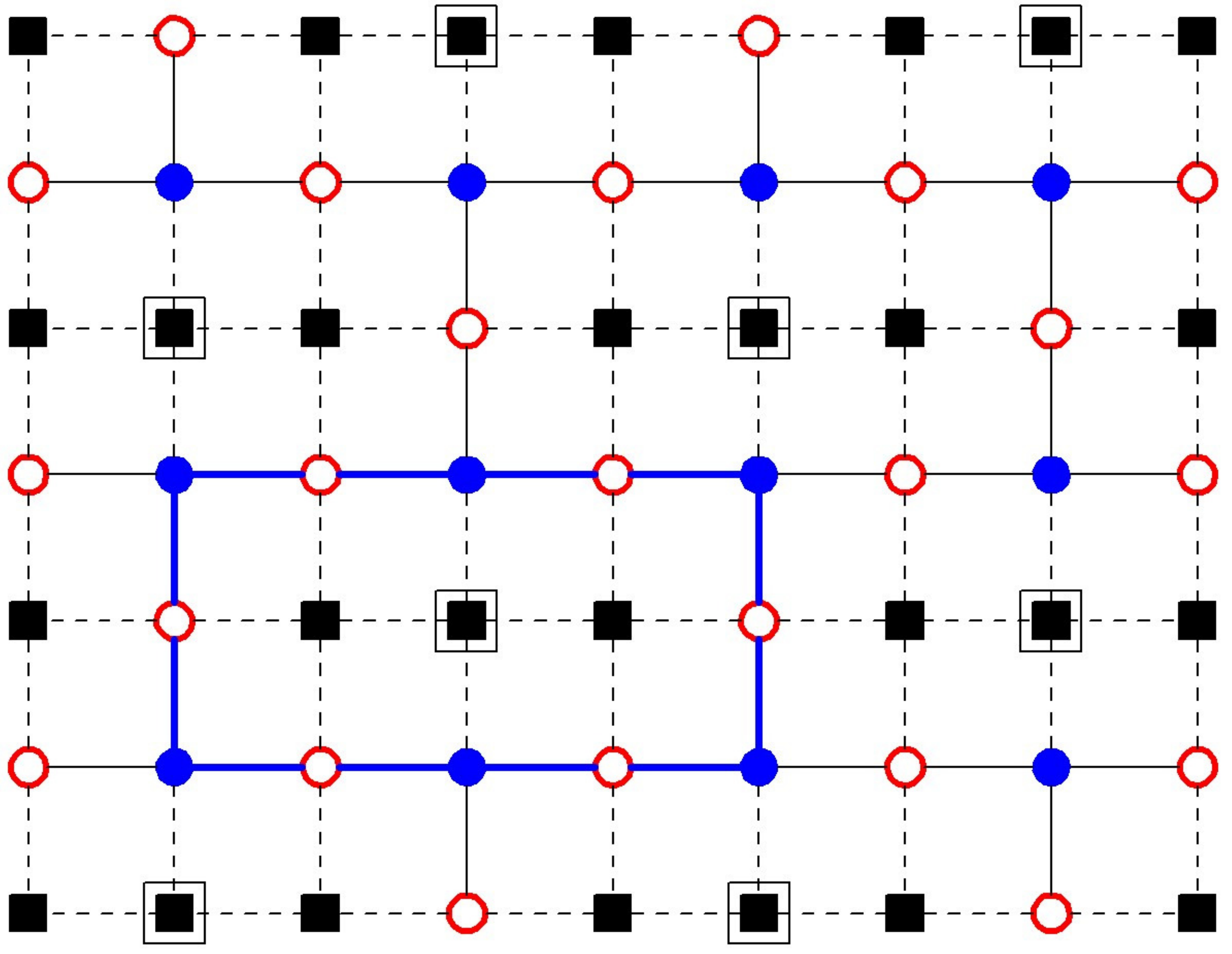}}
\caption{(Colour online) Schematic view of the transformed lattice. Open squares correspond to the atoms of the crystal
plane (1, 1, 1), closed squares - to the ghost lattice sites, open and closed circles - 
to $\alpha$ and $\gamma$-sites of the lattice model, respectively.} \label{fig2}
\end{figure}

It should be noted that the maximum number of nearest neighbours for the $\alpha$ and $\gamma$ 
sublattices are different;
i.e., depending on its position, the considered lattice site has a different number of the nearest
neighboring nodes. To simplify the program code of the simulation procedure, the lattice system can 
be supplemented with ghost sites, as shown in figure~\ref{fig2}. 
Ghost lattice sites are represented in this figure by closed squares. 
In this case, some of the introduced ghost sites coincide in position with the 
atoms of the crystal plane, which is reflected by the position of the closed squares 
inside the open ones, representing the atoms of the plane (figure~\ref{fig1}).

The introduction of ghost sites makes it possible to transform the original 
lattice structure into the ``square'' lattice, in which each lattice site has 4 nearest 
neighbors. In this case, any site of the sublattice $\alpha$ will have 2 fictitious 
sites among its nearest neighbors, and a site of the sublattice $\gamma$ 
will have only one neighboring ghost site.

Thus, each node of the modelled system $m(i,j)$ can be in one of 3 possible states:
$m(i,j)=1$ corresponds to a site $(i,j)$ occupied by the particle, $m(i,j)=0$ --- to 
a vacant lattice site, $m(i,j)=-1$ --- to a ghost lattice site. Obviously, during the 
simulation, the state of the ghost lattice sites does not change. In case of a 
random choice of the ghost lattice node, the selection procedure is not taken 
into account and a new node is selected.

If the change in the state of the lattice site leads to a decrease in the energy 
of the system ($\delta E_N\leqslant0$), then the new configuration is accepted. If the energy of the system
increases ($\delta E_N>0$), then the change in the state of the site is accepted with the probability
\begin{equation}
W=\exp(-\beta\delta E_N), \quad 
\beta=(k_{\text B}T)^{-1},
\label{eq15}    
\end{equation}
where $T$  is the temperature; $k_{\text B}$  is the Boltzmann constant.

In the latter case, to accept or reject the proposed change in the state of the lattice site, 
a random number $W_r$ is generated from the interval $[0;1]$, and if $W\gg W_r$, then the new 
configuration is accepted. Otherwise, the system reverts to its previous state. 
Repetition of the described procedure of $n$ times, where $n$ is equal to the number of 
particles on the lattice, forms one step of the Monte Carlo algorithm (MC step).

In this work, the modelled system contains $128\times128=2^{14}$ lattice sites. 
Of these, $5\times2^{11}$ sites are lattice sites of the two-level system, and $3\times2^{11}$ 
are the ghost lattice nodes. The complete simulation procedure consists of 
$30 000$ MC steps. Since the modelling procedure begins with a random distribution of particles 
on the lattice, the first $10 000$ MC steps are assigned to the transition of the system 
from an arbitrary initial state to an equilibrium state and are not taken into account 
in further modelling. Starting from $10 001$ steps of the algorithm, the number of particles 
in the system, the number of particles on each of the sublattices, and the number of 
pairs of nearest neighboring sites filled with particles are determined. 
On the basis of these results, after the end of the modelling procedure, the average 
concentration of particles in the entire system, the concentration of particles on 
the sublattices, the order parameter of the system, its thermodynamic factor, 
and the probability of detecting the two nearest lattice sites with occupied particles.

To reduce the effect of the size of the system on the obtained values of the equilibrium 
characteristics, we used periodic boundary conditions.

\section{Quasi-chemical approximation}

Together with the investigated lattice system, a similar reference lattice system can be considered~\cite{epj2000,PLA,epj2003}. This reference system is determined by the one-particle average potentials
$\phi^{(k)}(n_i^{\xi})$ of the interaction of a particle ($n_i^{\xi}=1$) or vacancy ($n_i^{\xi}=0$) 
located at site $i$ of the {$\xi$}-sublattice with some site~$j$, which is a neighbor of order $k$
($\xi=\alpha,\gamma; k=1$ corresponds to nearest neighbors, etc.). 
The potential energy of the reference system can be as
\begin{equation}
U_N^{(0)}=\epsilon_{\alpha}\sum\limits_{\left<\alpha\right>}n_i+
\epsilon_{\gamma}\sum\limits_{\left<\gamma\right>}n_i+
\sum\limits_{i=1}^{N}\sum\limits_{k}z_{k}^{\xi}\phi^{(k)}(n_i^{\xi}),
\label{eq16}    
\end{equation}
where $z_{k}^{\xi}$ is the number of neighboring sites of order $k$ for the lattice site of the
$\xi$-sublattice. Summation over index $k$ implies summation over the so-called coordination shells, 
i.e., over complete sets of neighbors of order $k$.

All equilibrium characteristics of the original lattice system are determined by its partition function 
\begin{equation}
Q_N=\text{Sp}_{\{n_{1}^{\xi},...,n_{N}^{\xi}\}}\{\exp(-\beta U_N)\},
\label{eq17}    
\end{equation}
where $\text{Sp}$ denotes the summation over all possible particle distributions given by sets of
occupation numbers $\{n_{i}^{\xi}\}$, taking into account the normalization conditions
\begin{equation}
n_{\alpha}=\sum\limits_{i=1}^{N_{\alpha}}n_{i}^{\alpha}, \quad 
n_{\gamma}=\sum\limits_{i=1}^{N_{\gamma}}n_{i}^{\gamma}, \quad
n=n_{\alpha}+n_{\gamma}.
\label{eq18}    
\end{equation}

The partition function of the initial lattice system (\ref{eq17}) can be represented as
\begin{align}
Q_{N}&=Q_{N}^{(0)}\text{Sp}_{\{n_{1}^{\xi},...,n_{N}^{\xi}\}}
\left\{\exp\left[-\beta\left( U_N-U_{N}^{(0)}\right) \right]
\frac{\exp(-\beta U_{N}^{(0)})}{Q_{N}^{(0)}}\right\}
\nonumber\\
&=Q_{N}^{(0)}\left<\exp\left[-\beta\left( U_N-U_{N}^{(0)}\right) \right]\right>_{0}=Q_{N}^{(0)}Q_{N}^{(\text{d})},
\label{eq19}
\end{align}
where $Q_{N}^{(0)}$ and $Q_{N}^{(\text{d})}$  are the partition function of the reference system 
and the diagram part of the partition function, respectively:
\begin{equation}
Q_{N}^{(0)}=\text{Sp}_{\{n_{1}^{\xi},...,n_{N}^{\xi}\}}
\left\{\exp\left(-\beta U_{N}^{(0)}\right)\right\}, \quad 
Q_{N}^{(\text{d})}=\left<\exp\left[-\beta\left( U_N-U_{N}^{(0)}\right) \right]\right>_{0}.
\label{eq20}    
\end{equation}

Angular brackets with index 0 denote the averaging over the states of the reference system, the distribution
functions of which, due to the one-particle structure of its potential energy, can be written as the
product of the average concentrations of particles and vacancies.

The partition function of the reference lattice system can be calculated analytically
\begin{equation}
Q_{N}^{(0)}=\prod\limits_{\xi=\alpha\gamma}\left[\exp(-\beta\epsilon_{\xi}c_{\xi})
\prod\limits_{i=0}^{1}\left(\frac{Q_{i}^{\xi}}{c_{i}^{\xi}}\right)^{c_{i}^{\xi}}\right]^{N_{\xi}},
\quad 
Q_{n_{i}^{\xi}}^{\xi}=\exp\left[-\beta\left(\sum\limits_k \phi^{(k)}(n_{i}^{\xi})\right)\right].
\label{eq21}    
\end{equation}

On the other hand, the diagram part of the partition function can be represented as
\begin{equation}
Q_{N}^{(\text{d})}=\left<\prod\limits_{i=1}^{N}\prod\limits_{j=1}^{N}\left[
1+f_{ij}^{(k)}(n_{i}^{\xi},n_{j}^{\zeta})
\right]\right>_0,
\label{eq22}    
\end{equation}
where $f_{ij}^{(k)}$  are the Mayer-like functions \cite{mayer}
\begin{equation}
f_{ij}^{(k)}(n_{i}^{\xi},n_{j}^{\zeta})=\exp\left\{-\beta\left(
J_{k}n_{i}^{\xi}n_{j}^{\zeta}-\phi^{(k)}(n_{i}^{\xi})-\phi^{(k)}(n_{j}^{\zeta})\right)\right\}-1,
\label{eq23}    
\end{equation}
where sites $i$ and $j$ are neighbors of order $k$; $J_k$ is the interaction energy of particles
occupying the $k$-th order neighbors; $\xi,\zeta=\alpha,\gamma$.

Equation~(\ref{eq22}) can be rewritten as
\begin{equation}
Q_{N}^{(\text{d})}=1+\left<\sum\limits_{j<i=1}^{N}f_{ij}^{(k)}\right>_{0}+
\left<\sum\limits_{j<i<k}^{N}f_{ij}^{(k)}f_{jk}^{(p)}\right>_{0}+
\left<\sum\limits_{j<i<k<l}^{N}f_{ij}^{(k)}f_{kl}^{(p)}\right>_{0}+...\, .
\label{eq24}    
\end{equation}
This series allows for a clear geometric interpretation. Each of its members is associated with 
the diagram (graph) consisting of the lines (link) corresponding to the functions 
$f_{ij}^{(k)}$ connecting the lattice nodes $i$ and $j$.
The type of each link is determined by the index $k$, i.e., the degree of proximity of nodes $i$ and $j$.

When averaging each diagram, the summation is performed over the lattice sites, leading to the 
appearance of the weight coefficient (weight) of the diagram, equal to the number of ways of 
placing the diagram on the lattice. The weight of the diagram is determined by the geometric 
properties of both the diagram itself and the lattice on which the diagram is placed.

Equation~(\ref{eq19}) makes it possible to represent the free energy of the initial lattice per 
one lattice site as the sum of two terms 
\begin{equation}
F=F^{(0)}+F^{(\text{d})},
\label{eq25}    
\end{equation}
where $F^{(0)}$ and $F^{(\text{d})}$ are the reference and diagram parts of the free energy, respectively.

We have
\begin{align}
F^{(0)}&=-\frac{k_{\text{B}}T}{N}\ln Q_{N}^{(0)}=
\frac{1}{5}\left(3c_{1}^{\alpha}\epsilon_{\alpha}+2c_{1}^{\gamma}\epsilon_{\gamma}\right)+
\frac{3}{5}k_{\text{B}}T\sum\limits_{i=0}^{1}c_{i}^{\alpha}
\left(\ln{c_{i}^{\alpha}}-\sum\limits_{k}z_{k}^{\alpha}\ln{X_{i}^{\alpha (k)}}\right)
\nonumber\\
&+\frac{2}{5}k_{\text{B}}T\sum\limits_{i=0}^{1}c_{i}^{\gamma}
\left(\ln{c_{i}^{\gamma}}-\sum\limits_{k}z_{k}^{\gamma}\ln{X_{i}^{\gamma (k)}}\right),
\label{eq26}
\end{align}
where
\begin{equation}
X_{n_{i}}^{\xi (k)}=\exp\left(-\beta\phi^{(k)}(n_{i}^{\xi})\right),\quad \xi=\alpha,\gamma;
\label{eq27}    
\end{equation}
$z_{k}^{\xi}$ is the $k$-th coordination number for the sublattice $\xi$,
i.e., the number of neighboring sites of order $k$ on the sublattice $\xi$ 
($z_{1}^{\alpha}=z_{\alpha},\quad z_{1}^{\gamma}=z_{\gamma}$).

To calculate the diagram part of the free energy, the corresponding partition 
function can be represented as 
\begin{equation}
Q_{N}^{(\text{d})}=\exp\left[-\beta NF^{(\text{d})}\right]=1-\beta NF^{(\text{d})}+
\frac{1}{2!}\left(-\beta NF^{(\text{d})}\right)^2+... \,.
\label{eq28}    
\end{equation}
This makes it possible to determine the diagram part of the free energy $F^{(\text{d})}$ by selecting from
expansion (\ref{eq24}) diagrams with weight coefficients linear in $N$.

Renormalization using the mean potentials of the Mayer functions improves the convergence 
of the expansion of the free energy of the system. The mean potentials themselves are 
determined from the principle of minimum susceptibility of the 
free energy to their variations \cite{epj2000}
\begin{equation}
\left(\frac{\partial F}{\partial\phi^{(k)}(n_{i}^{\xi})}\right)_{T}=0.
\label{eq29}    
\end{equation}

In turn, the order parameter $\delta c$ (\ref{eq8}) can be found from the condition for 
the extremality of the free energy \cite{CMP,epj2003}
\begin{equation}
\left(\frac{\partial F(c,\delta c)}{\partial\delta c}\right)_{T,X_{i}^{\xi(k)}}=0,
\label{eq30}    
\end{equation}
which is equivalent to the condition of equality of chemical potentials on the sublattices.

Holding various parts of the expansion of the diagrammatic part of the free energy,
one can construct several approximate methods for calculating the free energy of a
lattice system. The quasi-chemical approximation (QChA) corresponds to taking into
account the contribution of only two-vertex graphs in the diagrammatic part of the free energy.

The use of the principle of minimum susceptibility leads to the following system of 
equations for determining the mean potentials of the nearest neighbors \cite{monogr}
\begin{equation}
X_{i}^{\alpha}=\sum\limits_{j=0}^{1}c_{j}^{\gamma}\frac{W_{ij}}{X_{i}^{\gamma}},\quad 
X_{i}^{\gamma}=\sum\limits_{j=0}^{1}c_{j}^{\alpha}\frac{W_{ij}}{X_{i}^{\alpha}},\quad
X_{n_{i}}^{\xi}=X_{n_{i}}^{\xi(1)},\quad
\xi=\alpha,\gamma,
\label{eq31}    
\end{equation}
where
\begin{equation}
W_{ij}=\exp\left(-\beta J_{1}n_{i}^{\alpha(\gamma)}n_{j}^{\gamma(\alpha)}\right).
\label{eq32}    
\end{equation}

If the order parameter of the system $\delta c$ is given, then the system of equations (\ref{eq31})
has an analytical solution, the substitution of which in relation (\ref{eq25}) 
leads to the following expression for the free energy of the lattice fluid in the 
quasi-chemical approximation,
\begin{align}
F^{(\text{QChA})}=F^{(0)}&=\frac{1}{5}\left(3c_{1}^{\alpha}\epsilon_{\alpha}+
2c_{1}^{\gamma}\epsilon_{\gamma}\right)+
\frac{3}{5}k_{\text{B}}T\sum\limits_{i=0}^{1}c_{i}^{\alpha}\ln{c_{i}^{\alpha}}+
\frac{2}{5}k_{\text{B}}T\sum\limits_{i=0}^{1}c_{i}^{\gamma}\ln{c_{i}^{\gamma}}
\nonumber\\
&+\frac{6}{5}\left(c_{1}^{\alpha}\ln\eta_{\alpha}+c_{1}^{\gamma}\ln\eta_{\gamma}
+\ln{X_{0}^{\alpha}X_{0}^{\gamma}}\right),
\label{eq33}
\end{align}
where
\begin{align}
\eta_{\alpha(\gamma)}&=-\frac{c_{1}^{\gamma(\alpha)}-c_{0}^{\alpha(\gamma)}-W(c_{0}^{\alpha(\gamma)}-
c_{0}^{\gamma(\alpha)})}{2c_{0}^{\alpha(\gamma)}}+
\nonumber\\
&+\sqrt{\left( \frac{c_{1}^{\gamma(\alpha)}-c_{0}^{\alpha(\gamma)}-W(c_{0}^{\alpha(\gamma)}-
c_{0}^{\gamma(\alpha)})}{2c_{0}^{\alpha(\gamma)}}\right)^{2}+W\frac{c_{1}^{\alpha(\gamma)}}
{c_{0}^{\alpha(\gamma)}}},
\label{eq34}
\end{align}
\begin{equation}
X_{0}^{\alpha}X_{0}^{\gamma}=c_{0}^{\alpha}+\frac{c_{0}^{\alpha}}{\eta_{\alpha}}=
c_{0}^{\gamma}+\frac{c_{0}^{\gamma}}{\eta_{\gamma}},\quad
W=\exp(-\beta J_1).
\label{eq35}   
\end{equation}

Calculations show that, within the framework of the QChA, the range of the average potentials of the 
reference system coincides with the range of the interaction potential in the initial lattice system.
Therefore, in the case of the lattice system with the interaction of the nearest neighbors, the 
average potentials for the second and more distant neighbors will be equal to zero
\begin{equation}
X_{n_i}^{\xi (k)}=1,\quad k>1, \quad  \xi=\alpha,\gamma.
\label{eq36}   
\end{equation}

In \cite{monogr,epj2000} it was shown that when the average potentials are determined in accordance with
relation~(\ref{eq31}), the diagram part of the free energy $F^{(\text{d})}$ vanishes. Thus, in the framework of QChA, 
the free energy of the lattice model under study turns out to be equal to the 
free energy of the reference system~$F^{(0)}$.

\section{Diagram approximation}

The most obvious way to refine the results of QChA is to consider additional terms in the expansion 
of the diagram part of the free energy. The implicitly proposed strategy can be realized in the 
framework of the so-called diagram approximation (DA) \cite{PLA}.

The main idea of this approximation is that the average potentials of the reference system are taken to be 
equal to their values obtained within the framework of the quasi-chemical approximation considered 
above. In this case, for the diagrammatic part of the free energy, an additional assumption is 
put forward that the contribution to the free energy of the entire diagram series is proportional 
to the contribution of the simplest nonzero graph in the quasi-chemical approximation.

In accordance with the properties of QChA, such a graph will be the simplest ring graph, i.e., a closed
irreducible connected diagram containing the minimum possible number of links and vertices. In the case
under consideration, such a graph contains 12 vertices, 6 of which belong to the
$\alpha$-sublattice and 6 belong to the $\gamma$-sublattice and has a weight coefficient equal to 1/5. The
graphically indicated object is shown in figure~\ref{fig1}. 
Considering relations (\ref{eq31}), the expression for the diagrammatic part of the 
free energy can be written as \cite{monogr}
\begin{equation}
F^{(\text{d})}=-\frac{1}{5}\lambda d^{m},\quad d=(W-1)\frac{\sqrt{c_{0}^{\alpha}c_{1}^{\alpha}
c_{0}^{\gamma}c_{1}^{\gamma}}}{X_{0}^{\alpha}X_{1}^{\alpha}X_{0}^{\gamma}X_{1}^{\gamma}},
\label{eq37}   
\end{equation}
where $m$ is the number of vertices in the simplest ring diagram. 
In the case under consideration, $m=12$.

The proportionality coefficient $\lambda$ can be determined from the condition of equality 
of the critical parameter of the model, 
$J_{1}/k_{\text{B}}T_{c}$, in the diagram approximation 
to its value obtained during the simulation of the lattice fluid by the Monte Carlo method.

Thus, the final expression for the free energy takes the form
\begin{equation}
F^{(\text{DA})}=F^{(\text{QChA})}-\frac{1}{5}\lambda d^{m}.
\label{eq38}   
\end{equation}

The information about the free energy of the lattice fluid makes it possible to determine any of its
equilibrium properties. In particular, the chemical potential of the system $\mu$, the 
thermodynamic factor~$\chi^{T}$, and the probability of two nearest sites to be occupied by 
particles $P(1;1)$ are defined as
\begin{equation}
\beta\mu=\left(\frac{\partial(\beta F)}{\partial c}\right)_{T}, 
\label{eq39}   
\end{equation}
\begin{equation}
\chi^{T}=\frac{\partial(\mu\beta)}{\partial\ln{c}},
\label{eq40}   
\end{equation}
\begin{equation}
P(1;1)=\frac{2}{z_{k}}\left(\frac{\partial F}{\partial J_{k}}\right)_{T}.
\label{eq41}   
\end{equation}

\section{Results and discussion}

The fulfillment of the condition (\ref{eq13}) leads to the fact that in the case of the system with 
repulsion of the nearest neighbors, the number of 
shallower sites exceeds the number of deeper sites.

At the same time, the critical parameter of the system turns out to be equal to the 
critical parameter of the similar system with attraction between particles \cite{bsu21}
\begin{equation}
\beta_{c}J=\frac{J}{k_{\text{B}}T_{c}}=3.737.
\label{eq42}   
\end{equation}
\begin{figure}[!h]
	\centerline{\includegraphics[width=0.50\textwidth]{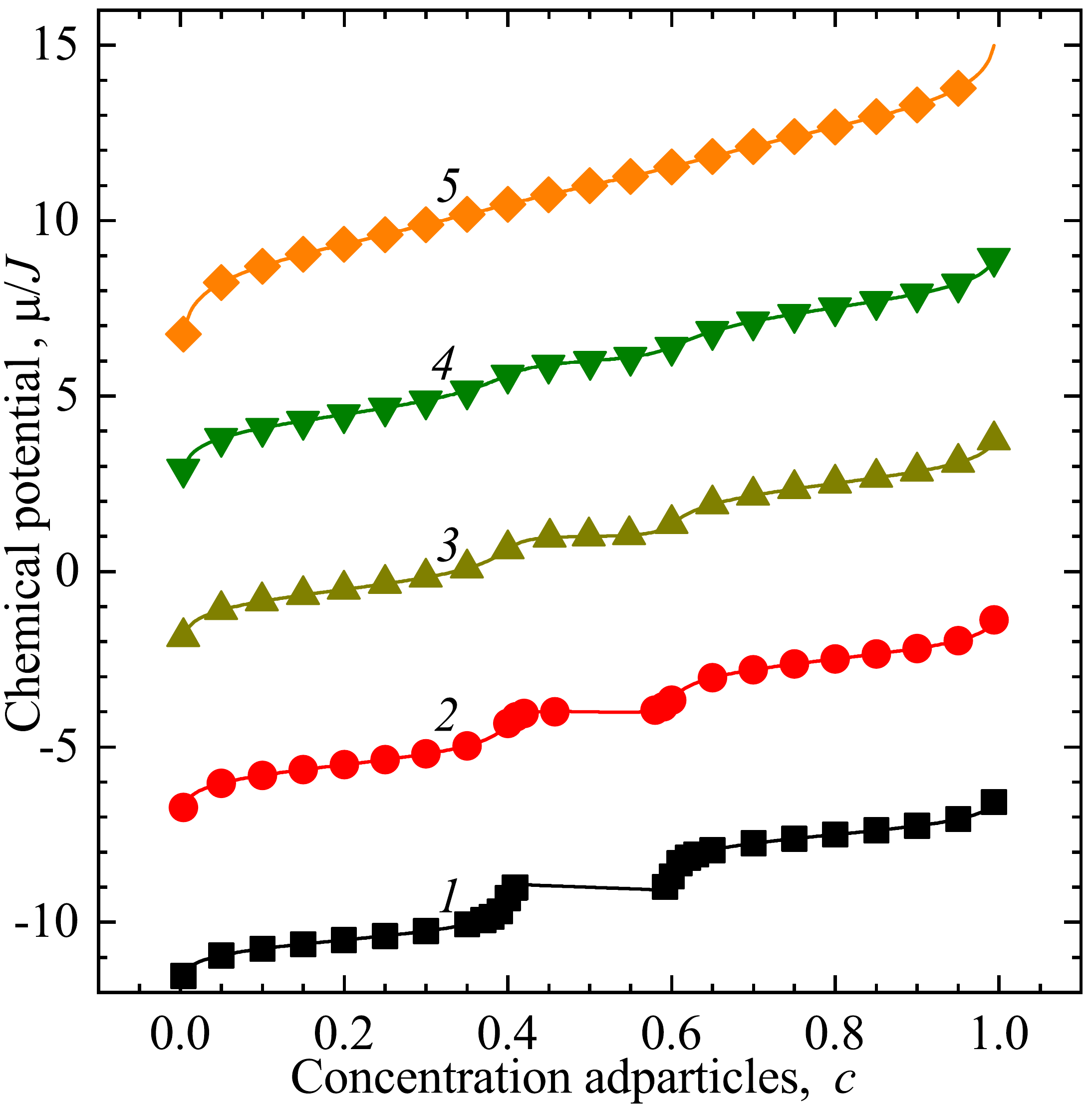}}
	\caption{(Colour online) The chemical potential (in units of the interaction energy of the nearest neighbors~$J$) versus
		concentration at $T/T_{c}=0.80$ (curves 1); 0.95 (2); 1.05 (3); 1.20 (4) and 2.00 (5). The solid lines
		represent the DA results, the full symbols are the MC simulation data. Each group of curves 
		is shifted down by 5 units with respect to the previous one for better visibility. 
		The unshifted curve~3 is characterized by $\mu/J=$ 1 at $c=0.5$, and this point is the same for 
		all the temperatures. Thus, the groups of curves 1 and 2 are shifted down from their true 
		position, while the groups 4 and 5 are shifted up.} \label{fig3}
\end{figure}
\begin{figure}[!hb]
	\centerline{\includegraphics[width=0.50\textwidth]{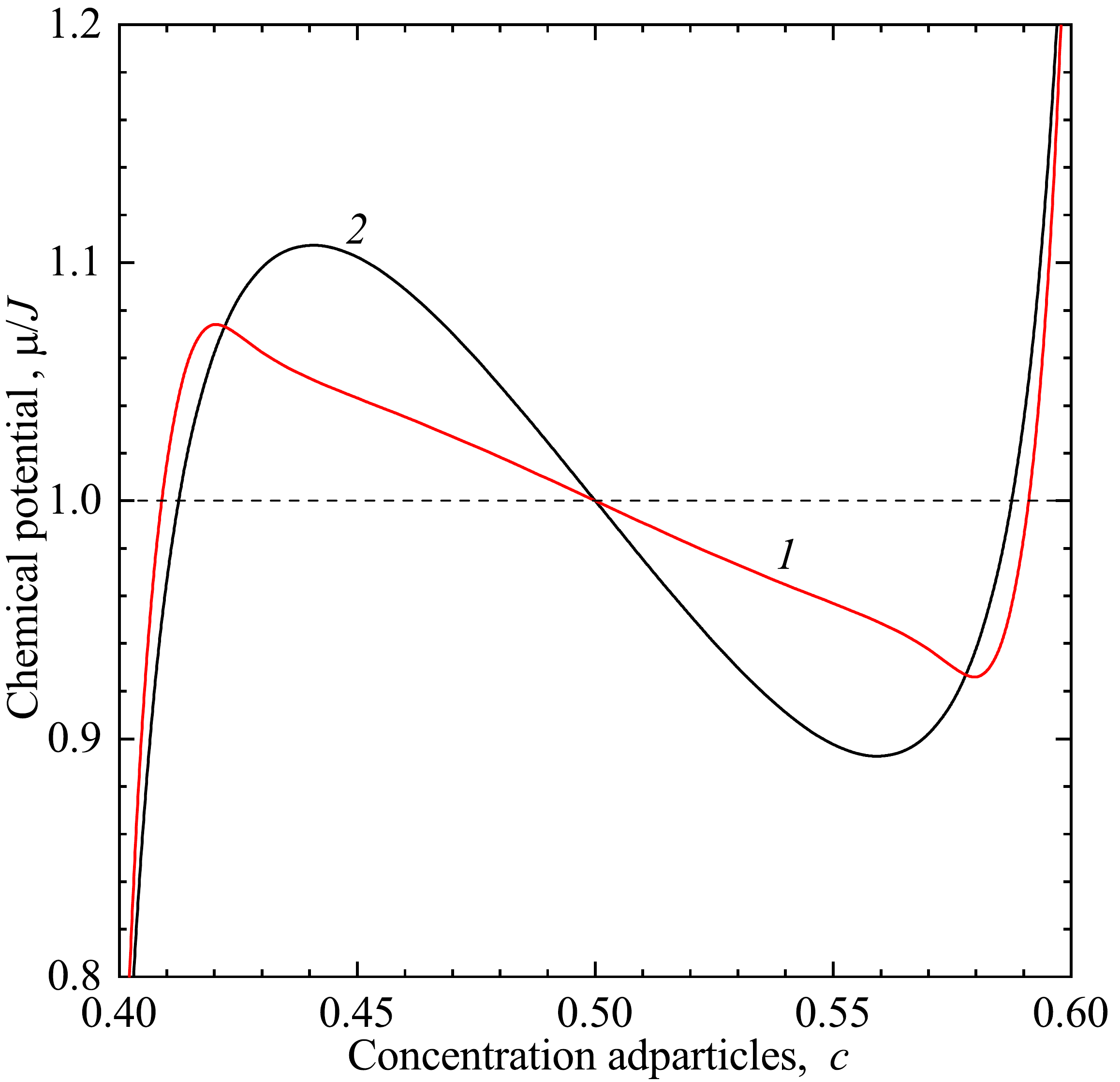}}
	\caption{(Colour online) Chemical potential isotherms at $T/T_{c}=0.80$. 
		Curve 1 represents the results of the quasi-chemical approximation, curve
		2 --- the diagram approximation.} \label{fig4}
\end{figure}

Comparison of the results of analytical calculations and modelling showed that the parameter 
$\lambda$ included in relation (\ref{eq38}), which ensures the equality of the critical parameters 
of the system found during its modelling and calculated within the framework of the diagram
approximation, is equal to 1.10308. 
It can be noted that its value differs very insignificantly 
from the analogous value obtained for a system with attraction 
between nearest neighbors \cite{bsu21}.
\begin{figure}[!ht]
	\centerline{\includegraphics[width=0.44\textwidth]{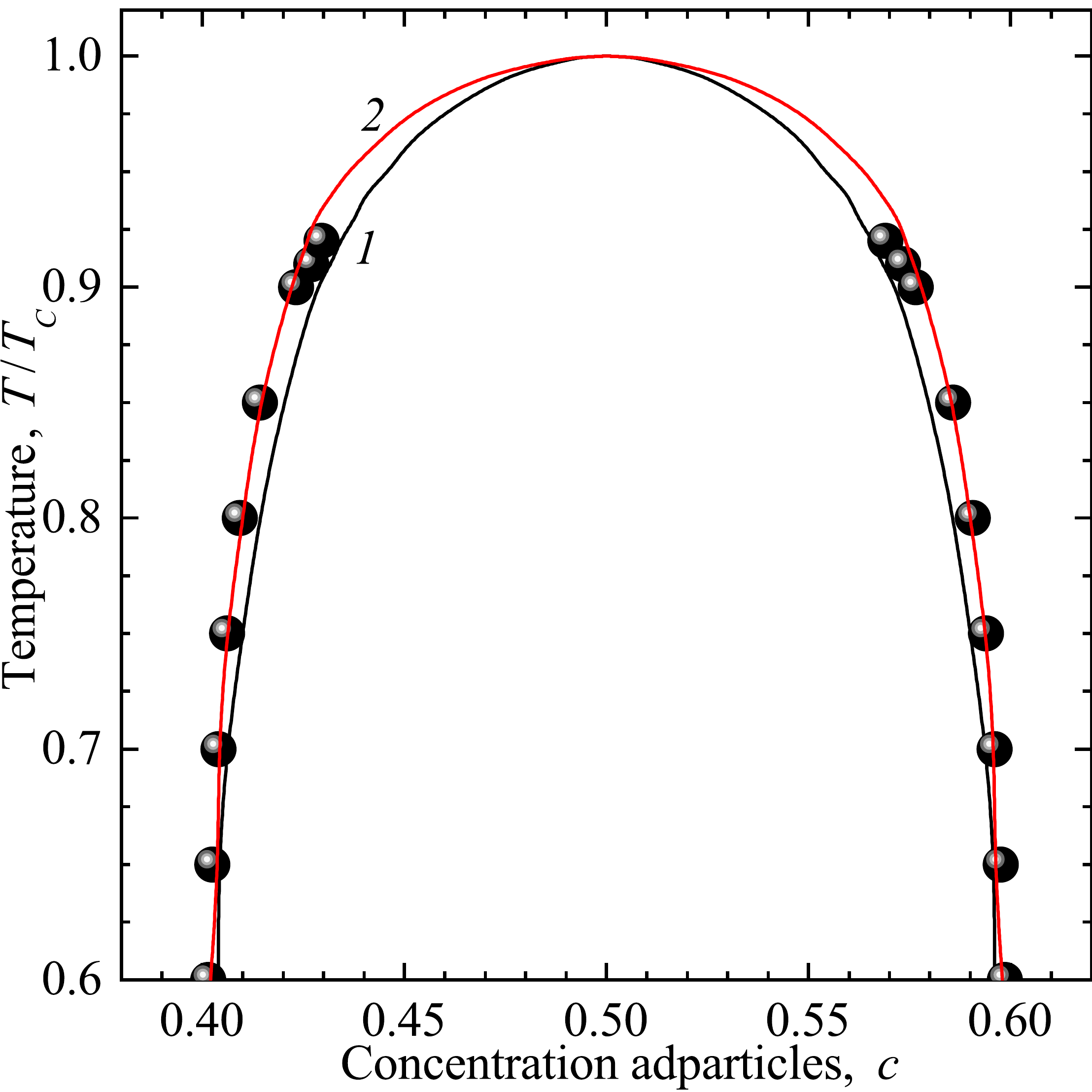}}
	\caption{(Colour online) Phase diagram of the lattice fluid with the repulsion of nearest neighbors 
		on a two-level lattice. Solid lines represent the results of the quasi-chemical 
		(curve~1) and the diagram (curve~2) approximations. 
		The full circles are the MC simulation results.} \label{fig5}
\end{figure}
\begin{figure}[!h]
	\centerline{\includegraphics[width=0.46\textwidth]{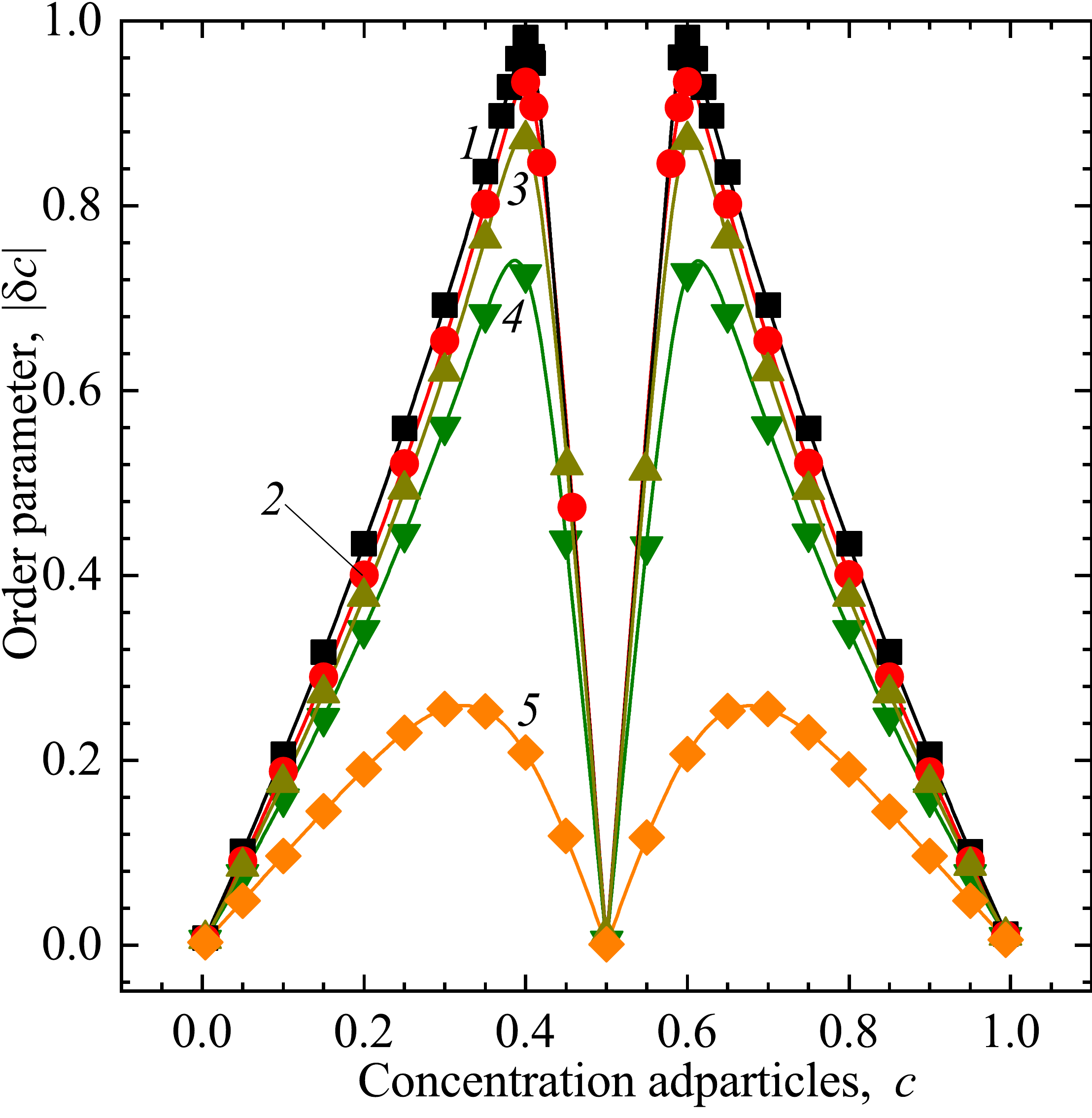}}
	\caption{(Colour online) The order parameter versus concentration for $T/T_{c}=0.80$ (curves 1);
		0.95 (2); 1.05 (3); 1.20 (4) and 2.00 (5). The solid lines represent the 
		diagram approximation results, the full circles are the MC simulation data.} \label{fig6}
\end{figure}

The isotherms of the chemical potential calculated for temperatures above and 
below the critical temperature $T_c$ are show in figure~\ref{fig3}. Figure~\ref{fig4} shows 
the isotherms of the chemical potential at a temperature of $0.8T_c$ determined within the 
framework of the quasi-chemical and diagram approximation.

It can be seen from the comparison (figure~\ref{fig3}) that the results of the diagram approximation are in good
agreement with the results of MC simulation in the entire range of the considered temperatures and
concentrations. As expected, in the two-phase region, the approximate approaches give a Van der Waals 
loop (see figure~\ref{fig4}), while the simulation of the system demonstrates the constancy 
of the chemical 
potential. This allows us to conclude that a first-order phase transition is observed in the 
case of a two-level lattice.

Maxwell's construction allows you to determine the points of phase transitions. 
The phase diagram of the system is shown in figure~\ref{fig5}.
The phase diagram turns out to be much narrower compared to the system with 
attractive interaction~\cite{bsu21}.
\begin{figure}[!h]
	\centerline{\includegraphics[width=0.47\textwidth]{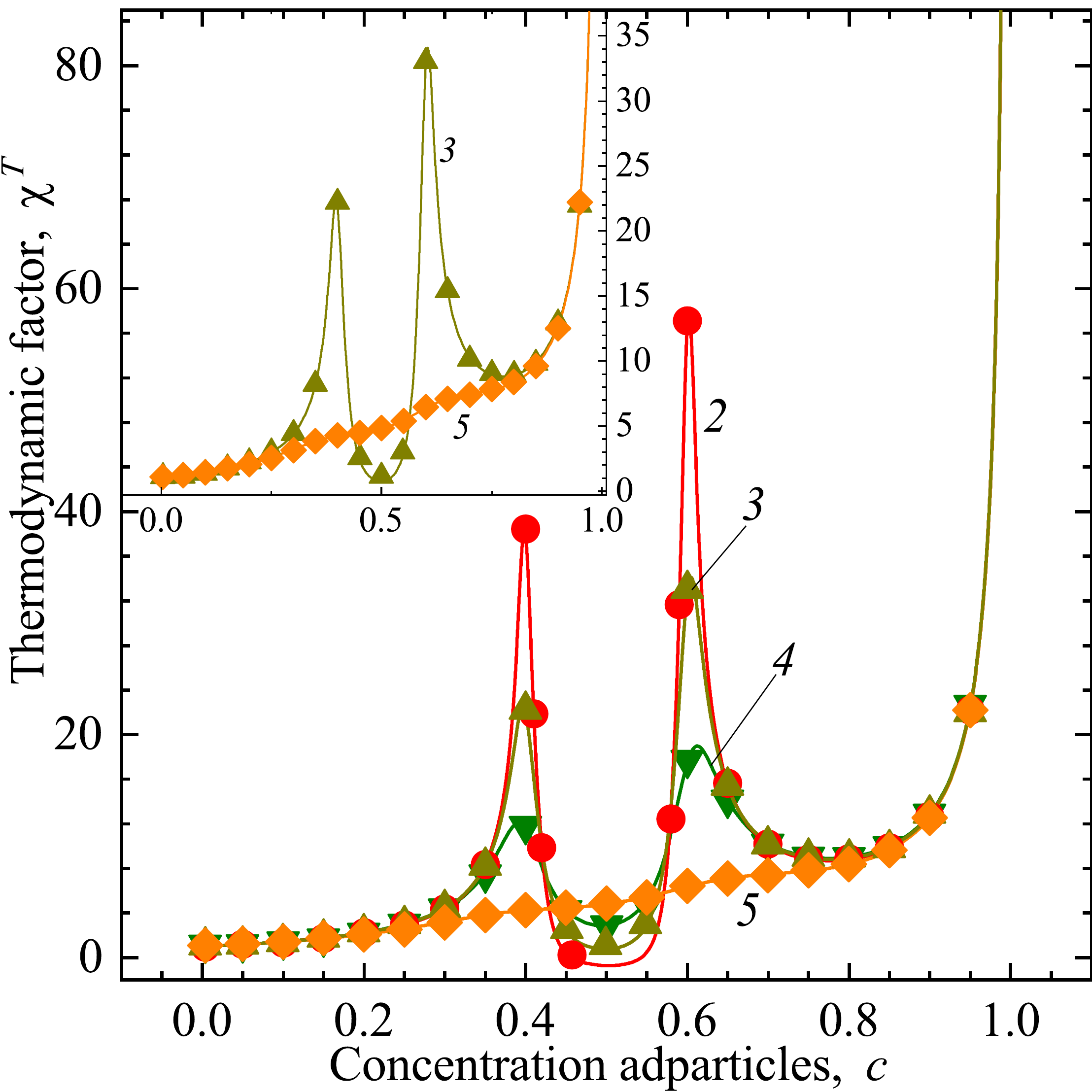}}
	\caption{(Colour online) The thermodynamic factor versus concentration at $T/T_{c}=0.95$ (curves 2); 
		1.05 (3); 1.20 (4)
		and 2.00 (5). The solid line represents the DA results, 
		the full circles are the MC simulation data.} \label{fig7}
\end{figure}
\begin{figure}[!h]
	\centerline{\includegraphics[width=0.47\textwidth]{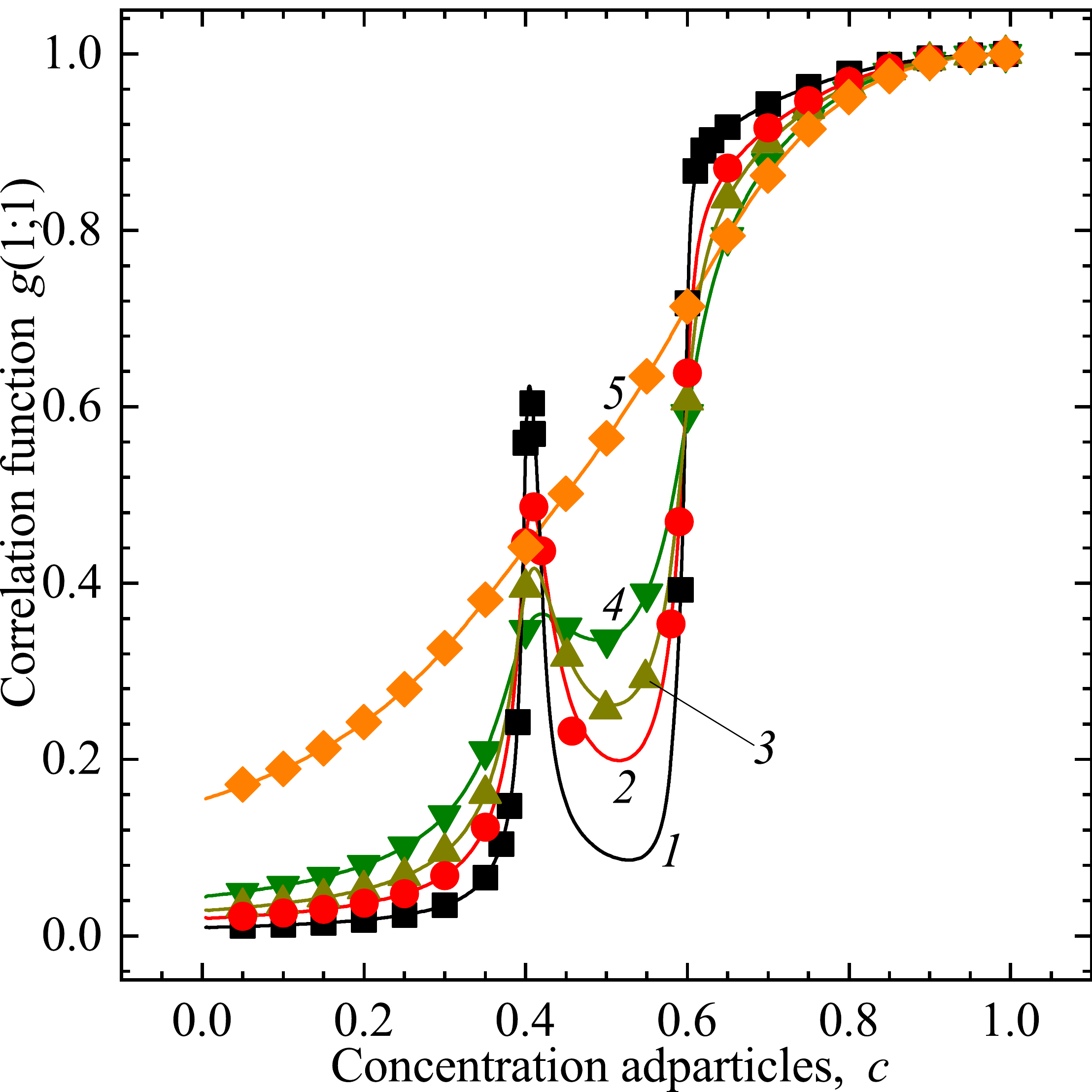}}
	\caption{(Colour online) The correlation functions for nearest neighbors versus concentration at $T/T_{c}=0.80$ 
		(curves 1); 0.95 (2); 1.05 (3); 1.20 (4) and 2.00 (5). 
		The solid lines represent the DA results, the full circles are the MC simulation data.} \label{fig8}
\end{figure}

In general, it can be noted that the thermodynamic characteristics do not qualitatively reveal 
the new properties in comparison with a similar system with 
the attraction of the nearest neighbors.

By contrast, the structural properties --- the order parameter (figure~\ref{fig6}), 
the thermodynamic factor (figure~\ref{fig7}) and the correlation function (figure~\ref{fig8}) ---
show the new features qualitatively.

For example, the order parameter is much larger than for the fluid with 
attraction between particles~\cite{bsu21}, and decreases monotonously with increasing 
temperature. In this case, in contrast to the one-level system~\cite{monogr}, 
the order parameter turns out to be nonzero even at temperatures above the critical one.

One can also see a sharp increase in the thermodynamic factor at particle concentrations near 0.4 and
0.6. This behavior of the thermodynamic factor is explained by the fact that at a sufficiently low
temperature and at concentrations less than 0.4, deeper $\gamma$-sites are predominantly filled. 
At $c=0.4$, the $\gamma$-sublattice is almost filled, and the transition of the particle to the
$\alpha$-sublattice is energetically unfavorable both because the $\alpha$-sites are 
shallower and because of the repulsion between the particles.

Thus, at a concentration of $c=0.4$, the mobility of particles sharply decreases. 
A similar situation is observed at a concentration of 0.6 with the only difference 
that, due to the interaction between particles at this concentration, the $\alpha$-sites
are predominantly filled, and the $\gamma$ sublattice is almost completely empty.

The features in the dependence of the correlation functions on the concentration shown in
figure~\ref{fig8} can be explained in a similar way. 
The sharp increase in the correlation function in the concentration range of 
0.4 and 0.6 for temperatures close to the critical one is due to an equally 
sharp decrease in the concentration of particles on one of the sublattices.

\section{Conclusion}

Comparison of the results of the developed approximate methods and simulation data shows that 
these results are in full qualitative and quantitative agreement in the entire range of temperature 
and concentration variation. 
This makes it possible to use, for example, the diagram approximation to study the thermodynamic 
and structural characteristics of the lattice fluid on a two-level lattice, 
since DA does not require powerful computers as compared with MC method.

The proposed quasi-chemical approximation refinement method is not the only possible one. 
Another strategy can be implemented within the framework 
of the so-called self-consistent diagram approximation. The main idea of the latter is to 
consider a reference system in which the range of the average potentials exceeds the range of 
the interaction potentials \cite{epj2000}. However, this approach leads to a significant 
complication of the obtained relations, which complicates their subsequent analysis.

It can also be noted that in the case of the system with repulsive interaction, the 
transition from a one-level to a two-level system changes the kind of phase transition in it.

To explain this fact, the isotherm of the chemical potential at low temperature (see curve 1 in figure~\ref{fig3}) 
must be considered together with the corresponding dependence of the order 
parameter versus concentration. 
The number of particles on the lattice increases monotonously with an increase in the chemical potential at low
temperatures and low concentrations. In this case, as follows from figure~\ref{fig6} (see curve 1), deep $\gamma$-sites 
are predominantly filled.
Obviously, such distribution of particles on the lattice is possible up to the concentration of 0.4. When this limit is
reached, all $\gamma$-sites will be occupied and additional particles can be placed on shallow $\alpha$-sites. In this case,
each particle on $\alpha$-sublattice will receive~2 nearest neighbors. 
Therefore, due to the interaction between particles, the concentration of particles increases with a jump to 0.6 
when the chemical potential reaches a critical value. 
This corresponds to the situation when all the $\alpha$-sites are occupied, 
and the $\gamma$-sites are vacant. There is a plateau on the isotherm of the chemical potential which 
corresponds to the phase transitions of the first order. 

Thus, the phase transition itself  corresponds to the transition between two different ordered states of the system. 
In the first of them, the deep $\gamma$-sublattice is predominantly occupied, and in the second, the shallow 
$\alpha$-sublattice is predominantly occupied.
This conclusion is fully confirmed by the analysis of snap-shots of the simulation procedure.
An increase in temperature reduces the ordering of the system. However, this ordering is essential at a 
sufficiently high temperature $2T_c$ (see curve 5 in figure~\ref{fig6}) and does not disappear even at extremely high 
temperatures ($10^{3}T_c$).

\section*{Acknowledgements}

The project is co-financed by the Polish National Agency for Academic Exchange
within ``Solidarity with scientists Initiative'' and European Union's Horizon-2020 research 
and innovation program under
the Marie Sklodowska-Curie grant agreement No 734276. 
I would like to thank Prof. Alina Ciach for comments on the manuscript.



\ukrainianpart

\title[Рівноважні властивості дворівневого ґраткового флюїду]%
{Рівноважні властивості ґраткового флюїду на двохрівневій ґратці з непрямокутною геометрією та відштовхуванням між найближчими сусідами%
}

\author[Я. Г. Грода]{Я. Г. Грода}
\address{Білоруський державний технологічний університет, вул. Свердлова 13a, 220006 Мінськ, Білорусь}

\makeukrtitle

\begin{abstract}
Досліджуються рівноважні властивості ґраткового	флюїду з відштовхуванням між найближчими сусідами на двохрівневій плоскій трикутній ґратці. Числові результати, отримані з аналітичних виразів, порівнюються з даними моделювання Монте-Карло. Показано, що запропоноване раніше діаграмне наближення дає можливість визначити рівноважні характеристики ґраткового флюїду з відштовхуванням між найближчими сусідами на двохрівневій ґратці з точністю, подібною до точності моделювання системи з використанням методу Монте-Карло, в цілому діапазоні термодинамічних параметрів. Виявлено, що на відміну від подібної однорівневої системи, ґратковий флюїд з відштовхуванням між найближчими сусідами зазнає фазового переходу першого роду. 
	\keywords ґратковий флюїд, дворівнева ґратка, квазіхімічне наближення, 
	діаграмне наближення, моделювання Монте-Карло
\end{abstract}
\end{document}